Sensitivity of Biomarkers to Changes in

Chemical Emissions in the Earth's Proterozoic Atmosphere

Grenfell, J.L.[1], Gebauer, S.[1], von Paris, P.[2], Godolt, M.[1],

Hedelt, P.[2*], Patzer, A.B.C.[1], Stracke, B.[2], and Rauer, H.[1,2]

(1) Zentrum für Astronomie und Astrophysik (ZAA)

Technische Universität Berlin (TUB)

Hardenbergstr. 36

10623 Berlin

Germany

(2) Institut für Planetenforschung (PF)

Deutsches Zentrum für Luft- und Raumfahrt (DLR)

Rutherford Str. 2

12489 Berlin

Germany

*Now at :

Laboratoire d'Astrophysique de Bordeaux (LAB)

Université de Bordeaux – CNRS

BP 89

33271  Floirac cedex

France

27 Pages, 5 Figures, 4 Tables.

Running Head: Atmosphere, Proterozoic, Biomarker

Editorial correspondence and proofs to:

Dr. John Lee Grenfell

Zentrum für Astronomie und Astrophysik (ZAA)

Technische Universität Berlin (TUB)

Hardenbergstr. 36

10623 Berlin

Germany

Email: lee.grenfell@dlr.de

*Abstract:* The search for life beyond the Solar System is a major activity in exoplanet science. However, even if an Earth-like planet were to be found, it is unlikely to be at a similar stage of evolution as the modern Earth. It is therefore of interest to investigate the sensitivity of biomarker signals for life as we know it for an Earth-like planet but at earlier stages of evolution. Here, we assess biomarkers i.e. species almost exclusively associated with life, in present-day and in 10% present atmospheric level oxygen atmospheres corresponding to the Earth's Proterozoic period. We investigate the impact of proposed enhanced microbial emissions of the biomarker nitrous oxide, which photolyses to form nitrogen oxides which can destroy the biomarker ozone. A major result of our work is regardless of the microbial activity producing nitrous oxide in the early anoxic ocean, a certain minimum ozone column can be expected to persist in Proterozoic-type atmospheres due to a stabilising feedback loop between ozone, nitrous oxide and the ultraviolet radiation field. Atmospheric nitrous oxide columns were enhanced by a factor of 51 for the Proterozoic "Canfield ocean" scenario with 100 times increased nitrous oxide surface emissions. In such a scenario nitrous oxide displays prominent spectral features, so may be more important as a biomarker than previously considered in such cases. The run with "Canfield ocean" nitrous oxide emissions enhanced by a factor of 100 also featured additional surface warming of 3.5K. Our results suggest that the Proterozoic ozone layer mostly survives the changes in composition which implies that it is indeed a good atmospheric biomarker.

*Key words: atmospheres, Proterozoic, photochemistry, biomarkers.*

**1. Introduction**

**Investigating atmospheric spectral signatures of species which potentially indicate life (biomarkers) on terrestrial exoplanets may be an attainable goal in the not-too-distant future. A central motivation is to assess how robust such signals are under varying atmospheric conditions and to differentiate signatures of life from the so-called "false positives" which mimic life.** Stars in the solar neighborhood are mainly young (e.g. Lopez-Santiago et al., 2004) hence, if they possess terrestrial planets, their age is expected to be comparable to the Archaean/Proterozoic periods in Earth's history assuming a similar evolution. Earlier works have assessed the responses of biomarker abundances in low oxygen ($O_2$) Earth-like atmospheres (Segura et al., 2003). By extending our results to exoplanets in their Habitable Zones (HZs), our approach considers planetary conditions with a similar history, photochemistry, biospheric input and development as in Early Earth's history.

Ozone ($O_3$) is considered to be a good biomarker for an Earth-like planet in the HZ because it is mostly formed from biogenic $O_2$ and it persists in the atmosphere even at very low $O_2$ levels (Kasting and Donahue, 1980; Segura et al., 2003). Previous studies have discussed the rise in atmospheric $O_2$ at the start of the Proterozoic (e.g. Catling 2005; Catling and Claire, 2005; Kump et al., 2001) representing their system as boxes (atmosphere, ocean, mantle etc.) interlinked with parameterised expressions for material fluxes. There are however some studies which applied coupled climate-photochemical column models. Some studies (e.g. Schindler and Kasting, 2000; Kaltenegger et al., 2007) presented theoretical spectra spanning the Archaean to the Proterozoic. $O_3$ is rapidly removed via catalytic cycles involving oxides of e.g. nitrogen ($NO_x$), chlorine ($ClO_x$), and hydrogen ($HO_x$) (Crutzen, 1970; World Meteorological Organization, 1995). The Proterozoic Earth likely featured enhanced volcanic and lightning emissions (Navarro-Gonzalez et al., 1998; Mvondo et al., 2001) which delivered $NO_x$ into the atmosphere hence influenced the presence of $O_3$. **During the Proterozoic, $NO_x$ recycling (i.e. conversion into its** **main atmospheric reservoir, nitric acid ($HNO_3$) and back again) may have been faster than today due to stronger solar UV fluxes, which stimulates both $HNO_3$ production via $NO_2+OH+M \rightarrow HNO_3+M$, as well as photolytic loss of $HNO_3$ back into $NO_x$.** Detailed modelling is thus required to address the non-linear photochemical responses of $O_3$ to these atmospheric conditions.

Concentrations of methane (Pavlov et al., 2003) and molecular hydrogen (Tian et al., 2005; Catling, 2006) - which can affect not only temperature, but also influence the abundance of $HO_x$ hence (via established catalytic cycles) the biomarker $O_3$ - were probably higher in the Proterozoic than today. Consequently, another important aspect of investigation is how such changes impact the significance of $O_3$ as a biomarker. In a baseline study, Segura et al. (2003) investigated $O_3$ responses to lowering abundances of $O_2$ using fixed, modern-day concentrations of the long-term source gases: $H_2$, CO, $CH_3Cl$, and $N_2O$ at the surface. Our work builds on their study by additionally varying surface-emitted species under Proterozoic Earth conditions, i.e. with low $O_2$ abundances and a faint young Sun with reduced solar luminosity based on Gough (1981) as discussed e.g. in Ribas et al. (2005). Some of our imposed composition increases (e.g. a doubling in the methane flux) are rather moderate compared with some estimates for the Proterozoic (e.g. Pavlov et al., 2003) but are nevertheless within the uncertainty range and remain within conditions for which our model has been validated.

Another biomarker molecule, nitrous oxide ($N_2O$) is usually associated almost exclusively with microbial life on the Earth. However, spectral measurements are challenging due to weak and narrow absorption bands for modern atmospheric concentrations. In the Proterozoic period, emissions of $N_2O$ were probably higher than today (Buick, 2007) associated with strong microbial activity in the early anoxic ocean (the so-called "Canfield ocean"), but, on the other hand, possible UV feedbacks may have led to faster $N_2O$ removal via photolysis. Des Marais et al. (2002) and Traub and Jucks (2000) discuss the effect of increased $N_2O$ concentrations upon its spectral signature. Note that (unlike $O_3$) $N_2O$ is located mostly in the troposphere.

We investigate photochemical feedbacks between the two important biomarkers $O_3$ and $N_2O$ using a coupled radiative-photochemical model. Finally, we quantify differences in the spectral signatures (especially for $O_3$ and $N_2O$) between the modern Earth and its Proterozoic period by calculating theoretical line-by-line spectra based on the calculated model atmospheres.

## 2. About the Models and the Runs

### 2.1 Atmospheric convective-radiative-photochemical column model

The original 1D model is described in detail in Kasting et al. (1984), Segura et al. (2003), Grenfell et al. (2007)[a]. The model calculates chemical concentrations and temperature profiles from the planetary surface up to the lower mesosphere. Latest improvements in the model version used here are described in Rauer et al. (2010). The radiative part of the climate module uses a longwave radiation scheme, the Rapid Radiative Transfer Model (RRTM) based on Mlawer et al. (1997) and a shortwave code based on Toon et al. (1989) and Pavlov et al. (2000). Convective adjustment in the troposphere is based on either a moist or dry adiabat depending on whether the saturation pressure of water is exceeded (Ingersoll, 1969). The model employs the relative humidity distribution described in Manabe and Wetherald (1967). The chemistry scheme features 55 species with **212** reactions including the $HO_x$, $NO_x$, and $ClO_x$ families. Natural biogenic and source gas emissions were included.

The model calculates global-mean conditions. Oxidation of higher hydrocarbons ($C_xH_y$) was neglected for x≥2. Interactive biogeochemical cycles are neglected. To account for missing clouds, the surface albedo in the climate and chemistry modules was adjusted until the converged surface temperature, $T_0$, reaches the measured mean Earth value of 288.0 K. Clouds influence atmospheric spectra as suggested by Kaltenegger et al. (2007), who showed theoretical spectra for Earth-like atmospheres at various stages of development. Climatic effects of different cloud layers on e.g. $T_0$ are described elsewhere (Kitzmann et al., 2010) using a version of our model without chemistry. Combining the effect of clouds with atmospheric chemistry is a future task not included here.

Sulphur compounds may have had an important impact on the redox properties of the atmosphere during the Proterozoic (e.g. Kasting et al., 1989). Sulphur chemistry in the current model version is comparable to Segura et al. (2003) involving 16 sulphur-containing species participating in 62 sulphur reactions. Like the Segura study, we assumed $SO_2$ and $H_2S$ surface emissions of 58 and 3.3 Tg sulphur per year respectively.

In the modern Earth control run 1, surface emissions were adjusted to reproduce modern day ambient concentrations, an approach already adopted e.g. in Segura et al. (2003) and Grenfell et al. (2007)[a,b]. Table 1 shows the resulting fluxes adopted in the model. Differences between modelled and observed values in Table 1 are attributed to some missing physical processes e.g. microphysics, and missing chemical reactions in the model. Also, a global average model is not able to include the effects of e.g. non-linear latitudinal variations in chemistry and dynamics. Overall, however, Table 1 suggests a good correspondence of our model fluxes to observations.

Furthermore, in this work we employ the widely-used assumption that atmospheric $O_2$ forms an isoprofile – even in the case of low $O_2$ content at 10% Present Atmospheric Levels (PAL). We have confirmed this assumption by comparison with interactive $O_2$ calculations – further details will be provided by Gebauer et al. (in preparation).

**2.2 Theoretical spectra model** – output from the atmospheric model (temperature, concentration profiles) is passed as input into the radiative transfer spectral model, SQuIRRL (Schwarzschild Quadrature InfraRed Radiation Line-by-line, Schreier and Böttger, 2003). SQuIRRL is a line-by-line radiative transfer model, which uses molecular absorption line databases (like HITRAN or GEISA) for the calculation of absorption cross-sections and, optionally, continuum absorption corrections. Local thermodynamic equilibrium (LTE) is assumed. Using the absorption cross-sections from HITRAN 2004 (Rothman et al., 2005), SQuIRRL calculates the emission and absorption for a chosen line-of-sight through the atmosphere. A Planck function with the appropriate temperature of the model layer is used as an emission source. The absorption within the beam is calculated from the surface to a height of ~70 km, divided into approximately 1 km layers as in the photochemical part of the atmospheric column model. The atmosphere is assumed to be cloud and haze free without scattering, consistent with the treatment used in the atmospheric model applied in our work. The emission spectra are calculated using a pencil beam looking down on the atmosphere at a **viewing** angle of 38 degrees as adopted in the previous study of Segura et al. (2003).

Other spectral models in the literature (Selsis et al., 2002; Meadows and Crisp, 1996) comprise high-resolution (~1cm$^{-1}$ or better) line-by-line codes. The Selsis et al. scheme is based on MODTRAN. SQuIRRL has participated in an intercomparison (Melsheimer et al. 2005) of eight radiative transfer schemes, which suggested deviations of <10% for the major absorbers between the various models.

**2.3 About the runs**

**2.3.1 Modern Earth control run**

The modern Earth control run (run 1, Tables 1, 2) features adjusted surface emissions and an adjusted surface albedo set to 0.21, to reproduce modern-day conditions. The converged model calculated a surface temperature, $T_0$=288.0K, corresponding to the measured modern Earth mean value with surface concentrations: $CH_4$=1.6x10$^{-6}$, $H_2$=5.5x10$^{-7}$, $CO$=9x10$^{-8}$, $N_2O$=3x10$^{-7}$ and $CH_3Cl$=5x10$^{-10}$ volume mixing ratio (vmr), which correspond to the modern day (1990) atmosphere. With these values, our model calculated an $O_3$ column of 305 Dobson Units (DU) (1DU = 2.7x10$^{16}$ molecules cm$^{-2}$) comparable to observations of ~310DU (IPCC TAR 2007).

**2.3.2 Proterozoic low level $O_2$ Conditions (10% PAL)**

In general, $O_3$ is subject to a "self-repair" mechanism, whereby an initial lowering in $O_3$

leads to enhanced UV on lower atmospheric levels, which in turn photolyses more $O_2$, *creating* $O_3$. Further, an increase in surface emissions of $N_2O$ and $CH_3Cl$, is expected to release $NO_x$ and $ClO_x$ respectively into the stratosphere, which would likely lead to more $O_3$ *destruction* via the established catalytic cycles. On the other hand, more $NO_x$ in the troposphere (e.g. from enhanced lightning on the Early Earth) can lead to tropospheric $O_3$ *production* via the well-known smog mechanism (Haagen-Smit, 1951).

Increasing $CH_4$ – which may have been an important greenhouse gas during the Proterozoic (Pavlov et al. 2003) - is expected to cool the stratosphere hence can favour $O_3$ *production* due to a well-known temperature-dependent response in the Chapman cycle, via a slowing in the $O_3$ sink: $O+O_3 \rightarrow 2O_2$. Also, $CH_4$ strongly impacts $HO_x$ photochemistry (hence $O_3$): in the troposphere $CH_4$ is an important sink for OH, whereas in the stratosphere $CH_4$ can favour $HO_x$ production since it is oxidised to water, which photolyses into $HO_x$.

To investigate separately such complex mechanisms, we performed six runs corresponding to the Earth's Proterozoic period (runs 2 to 7, Table 2) with varying composition and having 0.943 times the modern incoming solar flux based on Gough (1981) which corresponds to conditions of about 0.7 Gyrs ago. Generally, where observational evidence is unclear (e.g. for $CH_3Cl$) rather moderate changes in composition were adopted in this work, so as not to be too far removed from an Earth-like composition where the model has been extensively tested. High $N_2O$ scenarios, were however, also performed, because there are indications of high emissions (Buick 2007) of this specie during the Proterozoic. Our chosen methane surface flux of two times the modern methane fluxes is based approximately on the study of Pavlov et al. (2003) who suggested an increased $CH_4$ flux by up to a factor ten times the modern value. Emissions of other species for the Proterozoic period were varied as shown in Table 1 as a sensitivity study. $CO_2$ was lowered to a pre-industrial concentration of $2.77 \times 10^{-4}$ vmr (IPCC Third Assessment Report (TAR), 2007). Start values of remaining long-lived gases were set to modern levels.

In summary, run 2 represents the Proterozoic control calculation (i.e. the same surface gas emissions as the modern Earth control run 1, but with 10% PAL $O_2$ and reduced solar luminosity).

Run 3 is as for run 2 but employed increased NO emissions from lightning by a factor of 10 (Navarro-Gonzalez et al., 1998; Mvondo et al., 2001). Run 4 adopts a 100 times increase in surface $N_2O$ emissions, based on the "Canfield" (anoxic) ocean scenario as discussed e.g. in Buick (2007). Run 5 adopts a doubling in $CH_4$ emissions (Pavlov et al., 2003). Run 6 features a 10-fold increase in $CH_3Cl$ surface emissions, which are rather uncertain for the modern Earth (Keppler et al., 2005). For run 7, $H_2$ surface fluxes were modified to +25Tg/year, assuming a higher $H_2$ outgassing rate during the Proterozoic era (Tian et al., 2005), compared with modern Earth conditions (having −5.5 $H_2$ Tg/year, where the minus sign denotes *loss* on the surface via deposition). **In runs (1-6) we adopted a constant dry deposition velocity for $H_2$ removal of $7.7 \times 10^{-4}$ cms$^{-1}$. In run 7 we adopted a constant, upwards $H_2$ flux**. The input parameters of all model runs are summarized in Table 2.

## 3. Results

Table 3 summarises column values in DU. Figures 1a and 1b show atmospheric profiles of the biomarkers $O_3$ and $N_2O$ respectively.

### 3.1 Ozone as a Proterozoic Biomarker

**Proterozoic control** - for the 10% PAL $O_2$ run 2 (Table 3), the resulting $O_3$ column is reduced by 36DU compared with the modern Earth value obtained in run 1, but it is still able to function as an important UV shield. For the 10% PAL $O_2$ runs, the $O_3$ peak profile (Figure 1a) is shifted downwards compared with the modern Earth profile run due to the well-known $O_3$ repair mechanism, as already described above **which led to a strong lowering in $O_3$ in the mid to upper stratosphere for the Proterozoic control run 2 compared with the modern day control run 1**. Like our work, Segura et al. (2003) also obtained similar $O_3$ values for their 10% PAL $O_2$, noting that the $O_3$ column remained quite robust to a lowering in $O_2$.

**Varying chemical composition** – only a small effect upon $O_3$ arose from increasing $CH_4$ (run 5, Figure 1a) and $H_2$ (run 7), since these two lines overlay the 10% PAL $O_2$ control (run 2) up to about 30km. Increasing $N_2O$ (run 4), which decomposes in the stratosphere to form $NO_x$, led to strong local $O_3$ loss in the lower stratosphere with a 32% reduction in the column compared to the Proterozoic control run 2 (see Table 3) due to the established $NO_x$ catalytic chemistry (Crutzen, 1970). Similarly, increasing $CH_3Cl$ (run 6), which photolyses to form $ClO_x$, also led to significant $O_3$ loss, but in the mid to upper stratosphere where the $ClO_x$ cycles are favoured and to a 41% reduction in the column compared with run 2 (Table 3). Increasing lightning (run 3) which leads to enhanced $NO_x$ in the troposphere (Figure 1a) resulted in enhanced $O_3$ here due to production from the smog mechanism (Haagen-Smit, 1951), leading to a 7% increase in the column value compared with run 2 . This supports the findings of Grenfell et al. (2006), who proposed that the smog mechanism could play an important role on the Early Earth.

### 3.2 Nitrous Oxide as a Proterozoic Biomarker

The results listed in Table 3 suggest a strong lowering in the $N_2O$ column (77 DU) in the Proterozoic reference run 2 compared with the modern day control result (233 DU). **Related to this response is the previously-mentioned strong reduction in $O_3$ in the mid to upper stratosphere since the associated increase in UV implied a stronger sink for $N_2O$**. To investigate the $N_2O$ response Table 4 summarises the main $N_2O$ atmospheric sinks in the stratosphere at 20km. Note that the atmospheric inorganic sources are negligibly slow and are therefore neglected. Table 4 suggests that the lower $N_2O$ abundances for the Proterozoic reference run 2 are associated with an increased photolytic loss rate compared with run 1.

Comparing run 2 with run 4 in Figure 1 b and Table 3, indicates that increasing the $N_2O$ flux by a factor of one hundred (run 4) leads to an increase in the atmospheric column density of $N_2O$ by a factor of about fifty-one. Thus, the enhanced molecular flux is not completely converted into a concentration increase of $N_2O$ e.g. because the higher UV in the early Earth's upper stratosphere favoured more photolytic loss as already discussed.

**3.2.1 $N_2O$ and $O_3$ feedback** – enhanced $N_2O$ surface fluxes, possibly associated with the Canfield ocean, imply increased $NO_x$, which is the product of $N_2O$ decomposition in the stratosphere. This reduces $O_3$ hence increases UV, implying a stronger $N_2O$ photolytic sink, opposing the original increase in $N_2O$ and acting as a negative feedback, as previously mentioned. To investigate this feedback in more detail, we performed a family of "Canfield Ocean Runs" (COR), all having similar conditions to run 4 but with varying $N_2O$ surface emissions by factors, e.g. F =1, 10, 50, 100, 500, 1000 and 10,000 relative to the modern Earth value. **To validate these high $N_2O$ levels in our RRTM climate scheme, we performed a detailed comparison of RRTM with the line-by-line model SQuIRRL (as described in 2.2) comparing infra-red fluxes at the top of the atmosphere. For F=500 the radiance fluxes in the relatively strong $N_2O$ absorbing bands between 10 and 25 microns agreed to within about 5%. Agreement was weaker at 4-5 microns where the two schemes differed by up to about 30% but absorption in this spectral range was overall very weak. The total integrated radiation fluxes differed by 1.4% (F=500) and 4.2% (F=10,000).**

Figure 2a summarises results for the Canfield Ocean runs. The continuous line shows the increase in atmospheric $N_2O$ column with increasing surface emissions. The short arrowed-feature shows the gradient of the line arising from the three runs with F=1, 10 and 50. Deviation away from this gradient at higher emissions implies a non-linear feedback between $N_2O$ surface emissions and its atmospheric concentration. Increasing emissions led to enhanced $NO_x$ (e.g. from $2.2 \times 10^{-10}$ vmr (F=1) up to $7.3 \times 10^{-8}$ vmr (F=1000) in the lower stratosphere at ~20km). This implies a lowering in the $O_3$ column (Figure 2a), therefore an increase in UV-B, hence a stronger $N_2O$ photolytic sink. Therefore, $N_2O$ concentrations in Figure 2a did not increase linearly with surface emissions. Interestingly, at high $N_2O$ surface emissions, the $O_3$ column converged - to a value of about 132 DU. In this regime, investigations imply that any further lowering in $O_3$, led to an increase in $O_2$ photolysis (via Chapman production), which balanced the catalytic loss from $NO_x$. Figure 2b shows the contribution of the Chapman- and smog-mechanisms to the $O_3$ production near 10km. Results suggest that Chapman is the dominant mechanism, although at very high $N_2O$ surface emissions, the smog mechanism catalysed by $NO_x$ could become significant. **To test whether the mechanism could operate with enhanced chlorine we repeated the F=500 and F=1000 cases with ten times enhanced surface natural $CH_3Cl$ emissions; we found that the effect upon $O_3$ was insifgnificant, suggesting that our proposed $N_2O$-$O_3$ coupled mechanism was still operating.**

Consequently our result implies that regardless of the level of microbial activity producing $N_2O$ in the early anoxic ocean, which is not well-determined, a certain minimum $O_3$ column can nevertheless be expected to persist in Proterozoic-type atmospheres due to a stabilising feedback loop between the two biomarker molecules $O_3$, $N_2O$ and the UV radiation field.

### 3.3 Methane ($CH_4$) abundance and temperature responses

**3.3.1 $CH_4$ abundance -** Figure 3a implies that increasing the $CH_4$ surface flux by a factor of 2 in run 5 compared with run 2 leads to an increase by a factor of 2.4 in surface $CH_4$ concentrations, which approach ~$10^{-5}$ vmr in run 5. Runs 2, 6 and 7 all featured rather abundant $CH_4$ compared with the modern day control run 1. The increases arose because such runs featured less OH - the main sink for $CH_4$ - by up to about a factor (2-3) compared with the modern Earth OH value run 1. Less OH was associated with the lowering in $O_2$ and with a colder, drier atmosphere in run 2 (Table 3), since OH is produced via $H_2O+O(^1D) \rightarrow 2OH$. The run with enhanced $N_2O$, yielding a decrease in $CH_4$, due to a doubling in tropospheric OH (near 10km) than the Proterozoic reference run 2, associated with the fast reaction: NO (via $N_2O$ decomposition) + $HO_2 \rightarrow NO_2$+OH.

**3.3.2 Temperature**

**Stratosphere** – the Proterozoic runs exhibited a cooling response (Figure 3b) of up to 50K in the stratosphere, related to the lowering in $O_3$, which is the main heater in this region. As a result, for most runs the temperature inversion at about 0.1 bar is generally weakened compared with the control run 1, and the stratosphere shrinks in the vertical, with the stratopause moving down from about 0.001 bar (in the modern Earth run) to about 0.005 bar (in the Proterozoic runs). Runs 4 (enhanced $N_2O$) and run 6 (enhanced $CH_3Cl$) which destroyed $O_3$ locally in the lower- and upper- stratosphere respectively, featured enhanced cooling in these regions.

**Troposphere** – a fainter solar luminosity in the 10% PAL $O_2$ scenario (run 2) led to a reduced surface temperature of 282K compared with 288K in the modern Earth control run 1. Of the Proterozoic runs, the scenario with increased $N_2O$ featured the warmest surface temperature ($T_o$=285K), whereas the other runs featured only small deviations, with all values in the range $T_o$=(281-282)K.

**3.4 Strong lightning activity as a false positive for life?**

**In Table 3, increased lightning (run 3) leads to an enhancement in the biomarker $O_3$ compared with the Proterozoic control case (run 2). Does lightning act as a "false positive" by "artificially" increasing the biomarker signal? An investigation of this $O_3$ increase revealed that lightning produced some extra $NO_x$ which yielded some smog $O_3$ formation i.e. strengthened the $O_3$ biomarker, therefore more UV-shielding. For a planet without life, hence probably with very low $O_2$, these responses would not occur, because the smog mechanism requires $O_2$ to form $O_3$ via $O_2+O+M \rightarrow O_3+M$. Lightning therefore, according to our results, is not a false positive but a favourable phenomenon, which enhances but does not mimic atmospheric life signals.**

**3.5 Theoretical Emission Spectra**

We calculated spectra for the mid-infrared (IR) region (2-20) μm using the SQuIRRL model described in section 2.2 at a spectral resolution, R = 100. Figure 4 compares the spectral region from (2.0 to 20) μm for the modern Earth control run 1 and the Proterozoic control run 2. The well-known fundamental bands for $H_2O$ (6.3μm), $O_3$ (9.6μm) and $CH_4$ (3.3μm), are all captured in Figure 4. The $CO_2$ emission feature near 15μm which indicates the atmospheric temperature inversion at the tropopause is observable only for the modern run, not for the Proterozoic run, where the inversion is weaker.

Figure 5 is as for Figure 4 but shows the "Canfield ocean" run with 100 times more $N_2O$ emissions. In this case, some of the $N_2O$ bands - especially the feature at 7.8μm - are clearly enhanced. Further testing has shown that the majority of absorption from these bands arose at an atmospheric height below about 10km in the troposphere.

Des Marais et al. (2002) and Traub and Jucks (2002) investigated the effect of increasing $N_2O$ in a theoretical spectral model and discussed under which conditions detections becomes possible. However, their imposed concentration changes were arbitrary. **The new contribution of our work is to model consistently $N_2O$ chemical concentrations based on indications of enhanced biogenic fluxes during the Proterozoic. Our result provides an example – based on our own planet – that $N_2O$ could be a potentially important biomarker. We remind however, that our study uses a cloud-free model, neglecting their influence on atmospheric spectra of Earth-like exoplanets (see e.g. Tinetti et al., 2006 and Kaltenegger et al., 2007).**

**4. Discussion and Conclusions**

**This work has presented a scenario based on Proterozoic Earth, namely with a Canfield ocean having high $N_2O$ emissions. Results suggest that $N_2O$ may be a more important biomarker than previously thought. Also, since $N_2O$ is an efficient greenhouse** **gas, this led to about 3.5K surface warming for a Canfield ocean scenario with one hundred times enhanced $N_2O$ surface emissions (run 4).**

**We have also presented a mechanism which suggests that $O_3$ can survive variations in source gas emissions for the Proterozoic at 10% PAL $O_2$. Feedbacks between $N_2O$, $O_3$ and UV imply that the $O_3$ column is expected to survive even high levels of $N_2O$ emissions from Canfield ocean scenarios, and its fundamental band is mostly unchanged in the emission spectrum. Therefore, $O_3$ is concluded to be a good biomarker for the Proterozoic period.**

**Acknowledgements**

We are grateful to James Kasting and to Antigona Segura for providing the initial code and for useful discussion. We thank two anonymous referees for their helpful comments. This research has been supported by the Helmholtz Gemeinschaft through the research alliance "Planetary Evolution and Life".

Tables

Table 1: Comparison of total surface emission fluxes (Tg ($10^{12}$g)/year) for observations, the modern-day control run by Segura et al. (2003) and for our modern control run. Values in brackets represent the natural (without manmade) component only. Negative values for $H_2$ imply loss from the atmosphere which corresponds to a net $H_2$ deposition velocity of $7.7 \times 10^{-4}$ cm s$^{-1}$.

| Scenario | CH$_4$ | N$_2$O | CH$_3$Cl | H$_2$ | CO | lightning NO | SO$_2$ | H$_2$S |
|---|---|---|---|---|---|---|---|---|
| Observations Modern Earth | 500 to 600# (150-270)* | 6.7 to 36.6# (4.6-16.8) | 2.9# (2.7) | -3* Ehhalt (1999) | 1060 (TAR 2007) | 12.2 Kaminski et al. (2008) | (2-50) Bluth et al. (1993) | >2.6 Kourtidis et al. (2008)## (uncertain) |
| Segura et al. (2003) control | 954 | 13.2 | 7.3 | -1.3 | 2350 | Not quoted | 58 | 3.3 |
| This work Control (run 1) | 531 | 4.3 | 3.4 | -5.5 | 1800 | 10.5 | 58 | 3.3 |

*Difficult to separate natural from manmade sources. #Taken from International Panel on Climate Change (IPCC 2001), IPCC TAR (2007) and references therein. ## sum of manmade traffic emissions for Europe and Asia.

Table 2: Overview of the various calculation scenarios. "Faint sun" denotes the factor by which the top of atmosphere incoming solar radiation flux is reduced compared with its modern value. "PAL" denotes Present Atmospheric Level. "Tg/year" denotes surface emissions in Tg/year. CO$_2$ was set to 355ppm in the control run and to a pre-industrial value of 277ppm in the other runs. All runs were calculated with an adjusted surface albedo of 0.21. *Negative H$_2$ values indicate net removal at the surface with a constant dry deposition velocity (see text), proportional to the H$_2$ concentration in the atmosphere, except in run 7 (high H$_2$ run) where a positive, constant, surface flux was prescribed.

| Run | Scenario | Faint Sun Factor | % PAL O$_2$ | CH$_4$ surface (Tg/yr) | CH$_3$Cl surface (Tg/yr) | N$_2$O surface (Tg/yr) | NO lightning (Tg/yr) | H$_{2*}$ surface (Tg/yr) |
|---|---|---|---|---|---|---|---|---|
| 1 | Modern Control | 1 | 100 | 531 | 3.4 | 4.3 | 10.5 | -5.5 |
| 2 | 10% O$_2$ Control | 0.943 | 10 | 531 | 3.4 | 4.3 | 10.5 | -11.4 |
| 3 | 10% O$_2$ x10 lightning | 0.943 | 10 | 531 | 3.4 | 4.3 | 105.0 | -5.0 |
| 4 | 10% O$_2$ x100 N$_2$O | 0.943 | 10 | 531 | 3.4 | 430 | 10.5 | -4.5 |
| 5 | 10% O$_2$ x2 CH$_4$ | 0.943 | 10 | 1062 | 3.4 | 4.3 | 10.5 | -25.3 |
| 6 | 10% O$_2$ x10 CH$_3$Cl | 0.943 | 10 | 531 | 34.0 | 4.3 | 10.5 | -7.1 |
| 7 | 10% O$_2$ high H$_2$ | 0.943 | 10 | 531 | 3.4 | 4.3 | 10.5 | +25.0 |

Table 3: Column densities (Dobson Units) of atmospheric species investigated.

| Run | Scenario | $O_3$ | $N_2O$ | $CH_4$ | $CH_3Cl$ | $H_2O$ |
|---|---|---|---|---|---|---|
| 1 | Modern Control | 305 | 233 | 1240 | 0.4 | $2.4 \times 10^6$ |
| 2 | 10%PAL $O_2$ control | 269 | 77 | 3260 | 1.0 | $1.4 \times 10^6$ |
| 3 | 10%PAL $O_2$ x10 lightning | 287 | 77 | 1185 | 0.3 | $1.4 \times 10^6$ |
| 4 | 10%PAL $O_2$ x100 $N_2O$ | 183 | 3924 | 1094 | 0.3 | $1.9 \times 10^6$ |
| 5 | 10%PAL $O_2$ x2 $CH_4$ | 264 | 75 | 7987 | 1.2 | $1.6 \times 10^6$ |
| 6 | 10%PAL $O_2$ x10 $CH_3Cl$ | 159 | 33 | 1670 | 4.9 | $1.4 \times 10^6$ |
| 7 | 10%PAL $O_2$ high $H_2$ | 266 | 75 | 3447 | 1.0 | $1.5 \times 10^6$ |

Table 4: Quantities relevant to the modeled $N_2O$ budget. Rates of atmospheric $N_2O$ sinks are output at 20km. Shown are chemical reaction rates (molecules cm$^{-3}$ s$^{-1}$) and column values (Dobson Units) integrated over the complete model atmosphere. Atmospheric (inorganic) sources are negligible and are not shown.

| Quantity | Modern Earth (run 1) | 10% PAL $O_2$ (run 2) |
|---|---|---|
| Rate $N_2O + h\nu$ | 43.9 | 616.2 |
| Rate $N_2O + O^1D$ | 44.6 | 21.1 |
| $O_3$ column (DU) | 305 | 269 |
| $N_2O$ column (DU) | 233 | 77 |

**Figure Captions**

Figure 1a: Atmospheric ozone ($O_3$) profiles. Key: Solid = modern-day control (run 1). Long-Dashed = 10% PAL $O_2$ control (run 2). Short-dashed = 10% PAL $O_2$ x10 lightning (run 3). Long-short dashed = 10% PAL $O_2$ x100 $N_2O$ (run 4). Dots = 10% PAL $O_2$ x2 $CH_4$ (run 5). Dot-dashed = 10% PAL $O_2$ x10 $CH_3Cl$ (run 6). Dot-dot dashed = 10% PAL $O_2$ x10 $H_2$ (run 7).

Figure 1b: As for Figure 1a but for nitrous oxide ($N_2O$).

Figure 2a: Processes related to ozone-nitrous oxide feedbacks. Y-axis shows atmospheric column amounts in Dobson Units (DU) and UV-B radiation ($Wm^{-2}$) in the upper troposphere (16km) . X-axis shows the factor by which the surface $N_2O$ emissions ($E_{N2O}$) are increased relative to the modern day control run (modern $E_{N2O}$= 4.3 Tg $N_2O$/year, Table 1) .

Figure 2b : Response of ozone photochemistry near 10km to changes in the surface $N_2O$ flux. "%smog" denotes the amount of ozone formed from the smog mechanism ($O_{3(smog)}$), "%Chapman" denotes the amount of ozone formed from the Chapman cycle ($O_{3(Chapman)}$) i.e. Chapman only, without catalytic loss cycles. These two quantities are based upon straightforward steady-state equations which assume $NO_2$ and $O^3P$ (calculated from the model) are in chemical balance.

Figure 3a: Atmospheric methane ($CH_4$) profiles. Key as for Figure 1.

Figure 3b: Atmospheric temperature (Kelvin) profiles. Key as for Figure 1.

Figure 4: Theoretical spectrum with resolution, R ($=\lambda/\Delta\lambda$) = 100, for modern Earth control run 1 and the Proterozoic reference run 2 for the (2-20) micron range.

Figure 5: As for Figure 4 but for modern Earth control run 1 and the Proterozoic run with x100 enhanced $N_2O$ run 4.

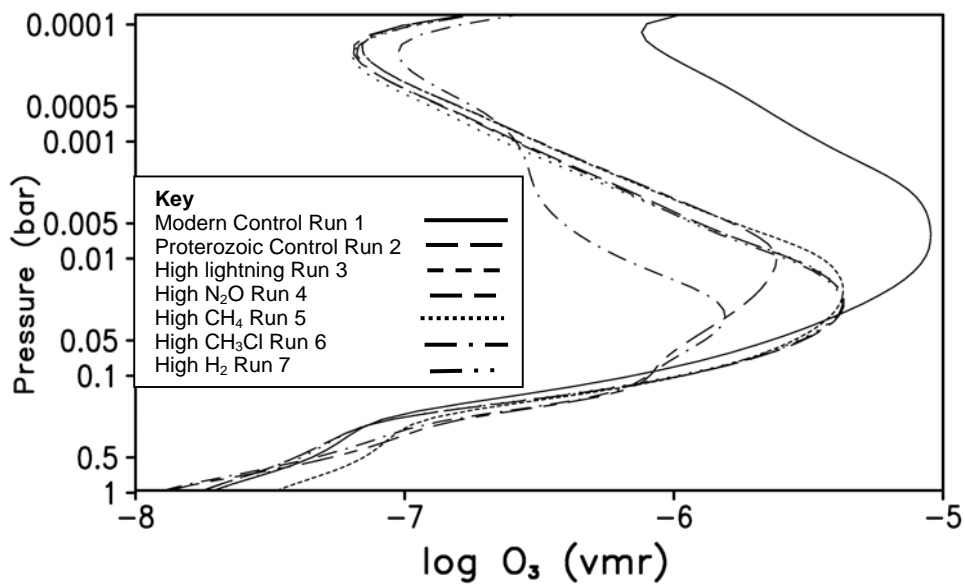

Figure 1a Ozone abundance

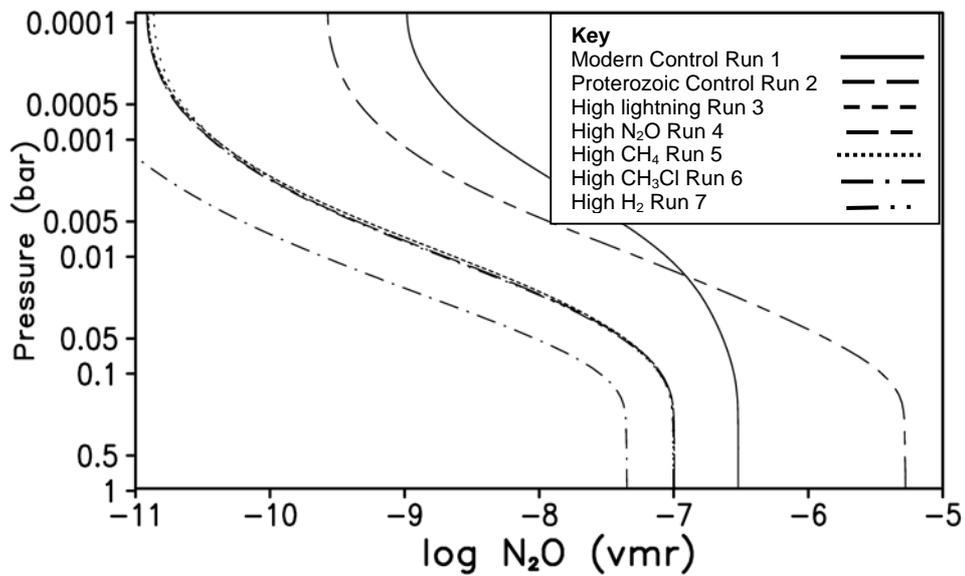

Figure 1b Nitrous oxide abundance

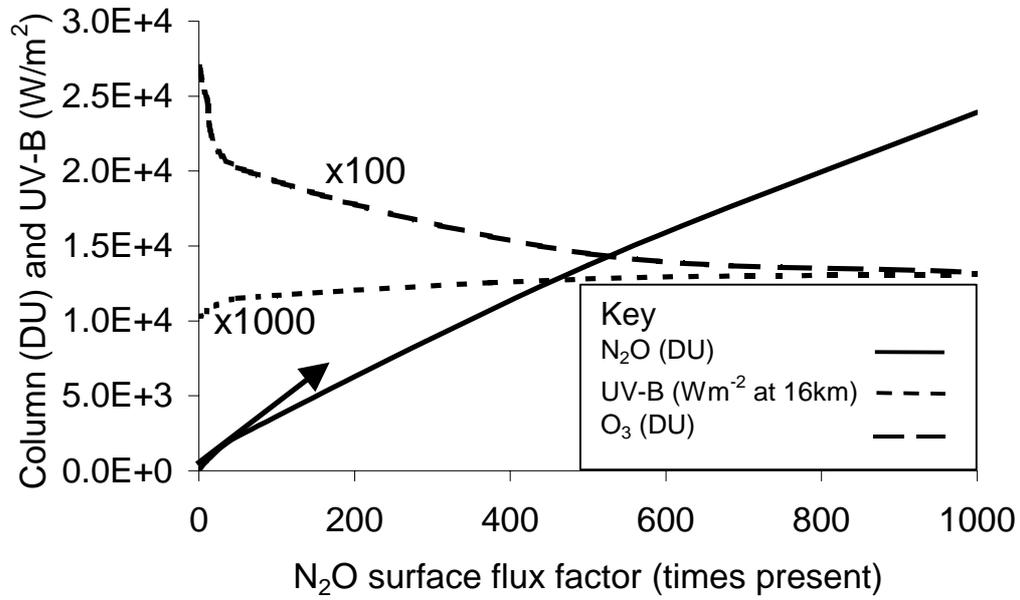

Figure 2a N$_2$O and O$_3$ feedbacks

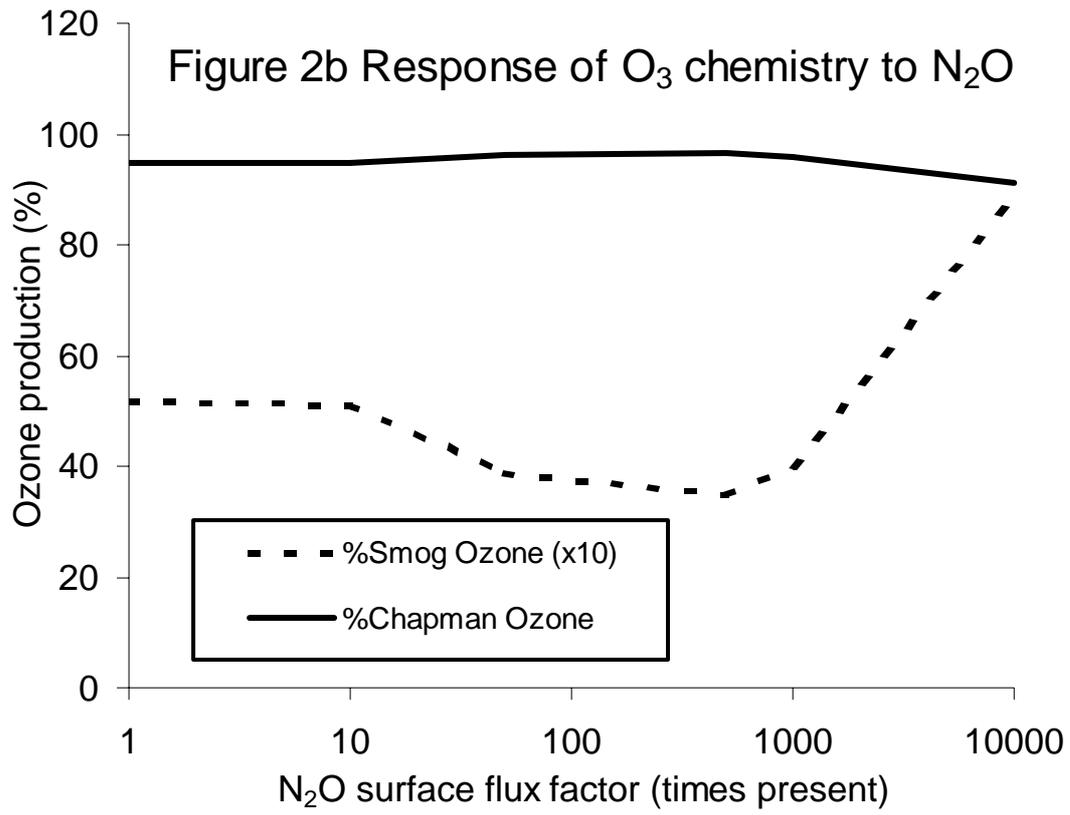

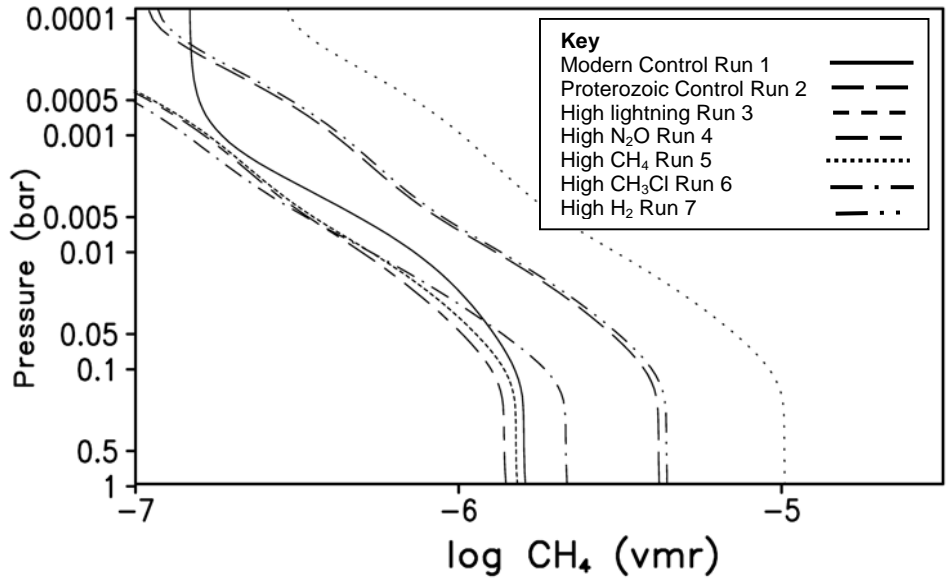

Figure 3a Methane abundance

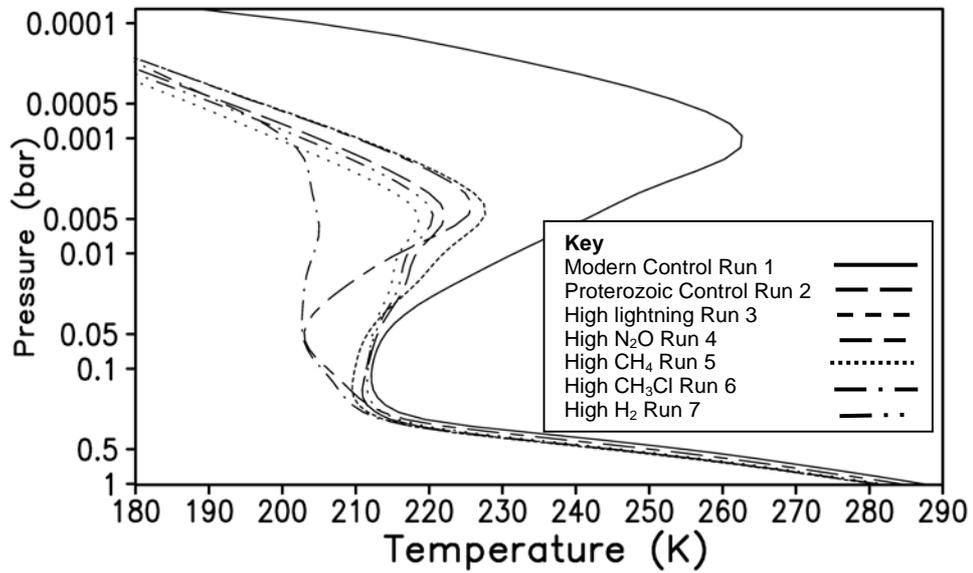

Figure 3b Temperature

**Figure 4: Synthetic spectra (R=100)**

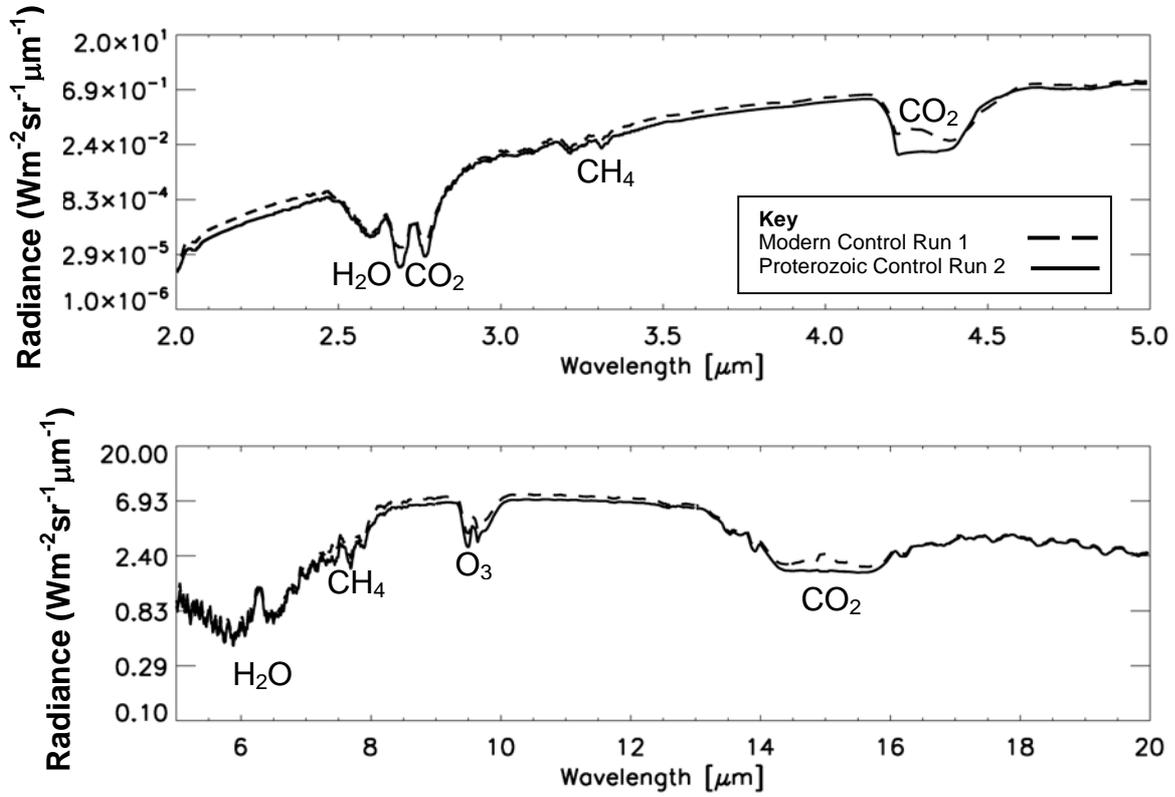

**Figure 5 Synthetic spectra (R=100)**

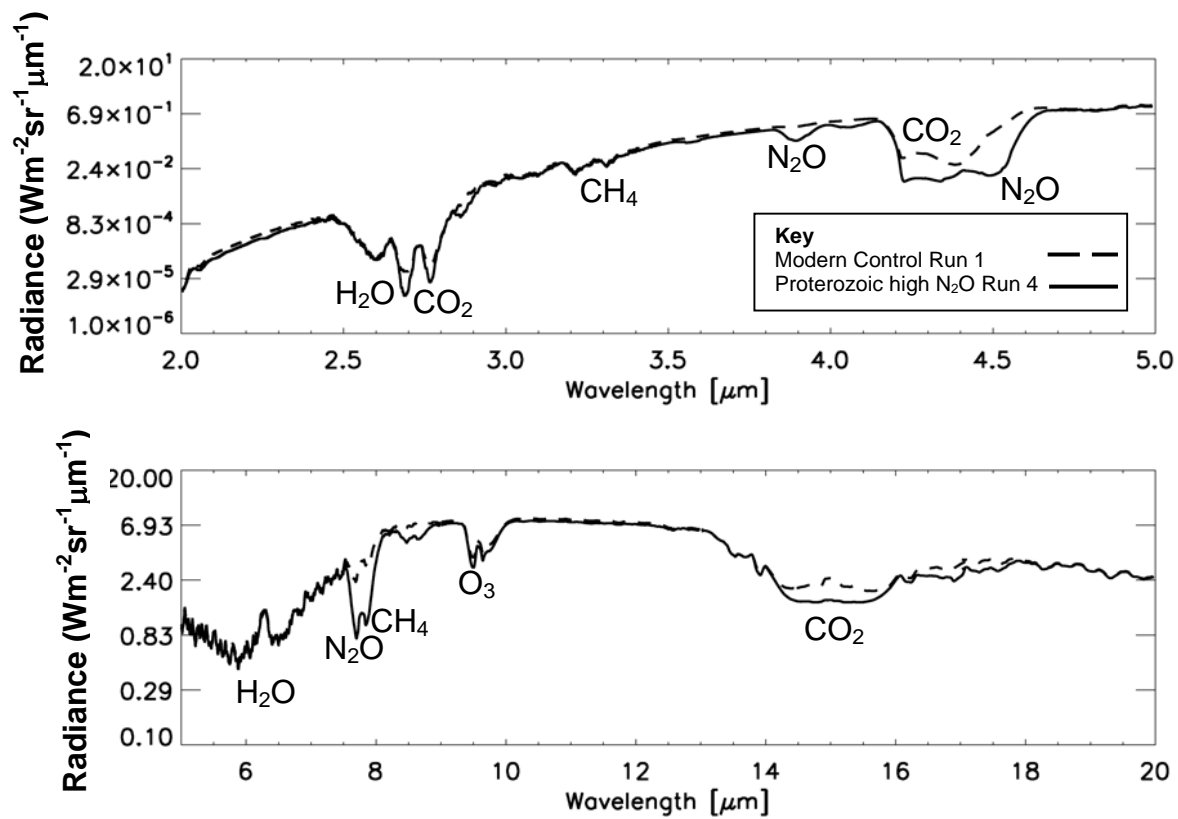